\documentclass[prl,onecolumn,12pts,preprintnumbers,amsmath,amssymb,superscriptaddress]{revtex4}

\usepackage{dcolumn}
\usepackage[pdftex]{graphicx}

\newcommand{\be}{\begin{equation}}
\newcommand{\ee}{\end{equation}}
\newcommand{\bea}{\begin{eqnarray}}
\newcommand{\eea}{\end{eqnarray}}
\newcommand{\bm}[1]{\mathbf{#1}}

\newcommand{\ra}{\rangle}

\newcommand{\lp}{\left(}
\newcommand{\rp}{\right)}

\def \cF{{\cal F}}
\def \cL{{\cal L}}

\newcommand{\ty}[1]{\mbox{\tiny #1}}

\begin{document}

\title{Moir\'{e} Butterflies}

\author{R. Bistritzer$^\star$ and A.H. MacDonald}
\affiliation{Department of Physics, The University of Texas at Austin, Austin Texas 78712 USA \\
 $^\star$rafib@physics.utexas.edu}

\date{\today}

\maketitle

{\bf The Hofstadter butterfly spectral patterns of
lattice electrons in an external magnetic field yield some of the most beguiling images in physics\cite{Harper,Azbel,hofstadter,streda,TKNN,AllanSlattice,Avron}.  In this
Letter we explore the magneto-electronic spectra of systems with  moir\'{e} spatial patterns\cite{Moire}, concentrating on the case of twisted bilayer graphene.  Because long-period spatial patterns are accurately formed at small
twist angles, fractal {\em butterfly} spectra and associated magneto-transport and magneto-mechanical anomalies emerge at
accessible magnetic field strengths.}

The fractal Hofstadter spectrum is a canonical example of electronic structure in a system with incommensurate length scales, and
has fascinated physicists and mathematicians for over a half a century.  The classic butterfly pattern is formed by the magnetic field dependent support of the eigenvalue spectrum of the
Schrodinger equation for a near-neighbor hopping model on a square lattice. 
Similar but distinct\cite{AllanHlattice} patterns describe the magneto-spectrum of any two-dimensional (2D) system of Bloch electrons.

Quite generally magneto-Bloch Hamiltonians are block diagonalizable
only when the magnetic flux through a
2D unit cell $\Phi$ is a rational multiple of the magnetic flux quantum $\Phi_0$. For  $\alpha\equiv\Phi_0/\Phi=p/q$
the spectrum consists of continuous subbands each containing an areal density of $B/q\Phi_0$, $q$ times smaller than
the usual semiclassical Landau level density.  Because the $x$ and $y$ components of cyclotron orbit centers are
canonically conjugate in a magnetic field\cite{LesHouches}, smearing the periodic potential over the magnetic length
scale $\ell = (\Phi_0/2 \pi B)^{1/2}$, the fractal pattern of gaps within Landau levels becomes visible only when $\alpha$ is not too much larger than one.
For atomic periodicity this condition is not met until the magnetic field strength exceeds laboratory scales by a factor
of about one thousand.  In moir\'{e} systems, however,  the pattern period is inversely proportional to the twist angle and
can easily exceed $\ell$.   Graphene moir\'{e} systems  realize Hofstadter physics at fields of a few Tesla, without recourse to the
difficult and potentially damaging photolithographic patterning used previously \cite{klitzing1} to
realize Hofstadter physics in the lab.

Because of the relatively weak forces between adjacent graphene layers,
double layer graphene systems with a variety of different stacking sequences
occur in bulk graphite\cite{rotatedGraphite}, epitaxially grown multi-layer graphene\cite{rotatedEpitaxial},
and in mechanically exfoliated multilayers \cite{twistExfoliated}.
Relative twists between layers can also be created by folding a single layer\cite{foldedBilayer,foldedBilayer2}.
The stacking arrangement in a two-layer system
can be characterized by the twist angle $\theta$, and by a relative translation $\bm{d}$.
Different electronic structure aspects of the twisted bilayer have captured
theoretical attention\cite{Santos,ShallcrossLong,localization,Morell,Mele,FLGus,moireBands}, and have already spurred
some experimental observations\cite{twistExfoliated,vanHove_Andrei}.

For $\theta$ smaller than roughly ten degrees, the low energy spectrum is faithfully described by a continuum model obtained via an
envelope function approximation\cite{moireBands,Santos}. For small twist angles, this model shows that it is
meaningful to describe the electronic structure in terms of Bloch bands for any $\theta$ despite the fact that the atomic
network is periodic only for a discrete set of angles.
The Bloch bands in this description are intimately related to the moir\'{e} pattern clearly observed in
scanning tunneling microscopy measurements\cite{rotatedEpitaxial}.
The moir\'{e} period for bilayer graphene is $D=a/[\sin(\theta/2)]$ where $a$ is graphene's lattice constant.
Because a translation of one layer with respect to the other only shifts the moir\'e pattern the
electronic structure is virtually independent of $\bm{d}$ \cite{moireBands} except at large commensurate twist angles.
In what follows, we therefore set $\bm{d}$ to zero.

The continuum limit of a $\pi$-band tight-binding model for the twisted bilayer yields a
transparent physical picture in which Dirac cones are coupled by a
position and sub-lattice dependent interlayer hopping $T(\bm{r})$ operator
that captures the local coordination of the twisted honeycomb lattices.
It is $T(\bm{r})$, and not a periodic potential, which is responsible for the moir\'{e} butterfly.
Because the moir\'{e} unit cell area  $\Omega_{\ty M} \propto \theta^{-2}$ (see Supplementary Information), the flux through a moir\'e unit cell
increases rapidly as the twist angle is reduced.  As in the periodic potential case, gaps open
within Landau levels for $\alpha \lesssim 1$. Because $B[T] \approx 4 \lp\theta^\circ\rp^2/\alpha$,
significant splitting of the isolated layers Dirac Landau levels appear already at low magnetic fields for small  $\theta$.

\begin{figure*}[h]
\includegraphics[width=0.8\linewidth]{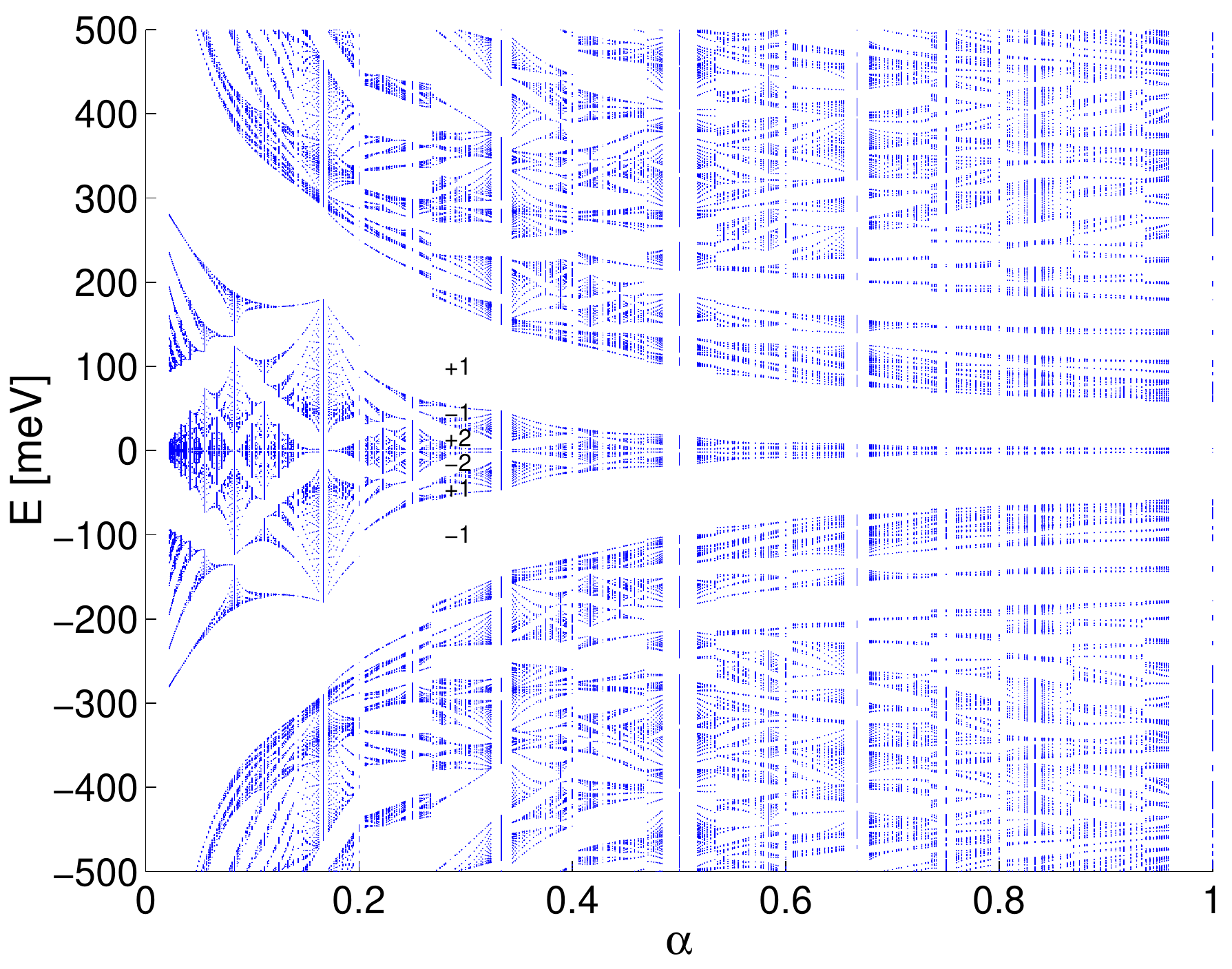}
\caption{{\bf Spectrum support.}
The support of the spectrum as a function of $\alpha$ for $\theta=2^\circ$ ($w=110$meV). The periodic inter-layer hopping amplitude
results in a Hofstadter-like sub-band structure. The integers denote the Hall conductivity associated with the larger energy gaps between $\nu=1$ and $\nu=-1$ for $\alpha \approx 0.3$.}
\label{fig:Esupport}
\end{figure*}

In the absence of inter-layer coupling, the spectrum consists of degenerate Dirac Landau-levels at
energies  $\pm \omega_c \sqrt{n}$, where $\omega_c=\sqrt{2}v/\ell$ is the cyclotron energy, and
$v$ is the graphene sheet Dirac velocity.  Because of the layer degeneracy,
gaps between Landau levels produce quantum Hall effects at odd, rather than half odd\cite{QH_graphene}, integer filling factors.  
(Spin and valley degeneracy are left implicit throughout this article.)
Coupling between the layers splits the Landau levels in both layers into $q$ sub-bands and
couples them together as illustrated in Fig.\ref{fig:Esupport} for the $\theta=2^\circ$ case.
It is clear that interlayer coupling at strong fields completely alters the spectrum.

We now briefly derive the equations we use to evaluate the moir\'{e} butterfly spectrum at
rational values of $\alpha$, present numerical results for a typical twist angle,
and discuss the magneto-transport and magneto-mechanical anomalies that they imply.

The low energy electronic structure of a twisted bilayer is well captured by the continuum model in which\cite{moireBands}
\be
H = \left(
      \begin{array}{cc}
        h(-\theta/2) & T(\bm{r}) \\
        T^\dagger(\bm{r}) & h(\theta/2) \\
      \end{array}
    \right)         \label{Hr}
\ee
where $h=i v\sigma \cdot \nabla$ with $\sigma=(\sigma_x,\sigma_y)$ being the sub-lattice Pauli matrices of the single-layer graphene Hamiltonian, and
\be
T(\bm{r}) = w \sum_j e^{-i\bm{q}_j \cdot \bm{r}} \; T_j        \label{Tr}
\ee
is the inter-layer hopping matrix. Here
\be
T_{1} =
\left(
  \begin{array}{cc}
    1 & 1 \\
    1 & 1 \\
  \end{array}
\right),
T_{2} =
\left(
  \begin{array}{cc}
    e^{-i\phi} & 1 \\
    e^{i\phi} & e^{-i\phi} \\
  \end{array}
\right),
T_{3} =
\left(
  \begin{array}{cc}
    e^{i\phi} & 1 \\
    e^{-i\phi} & e^{i\phi} \\
  \end{array}
\right),                \label{Tmatrices}
\ee
where $\phi=2\pi/3$, $\bm{q}_1=k_\theta(0,-1)$, $\bm{q}_2=k_\theta(\sqrt{3},1)/2$, $\bm{q}_3=k_\theta(-\sqrt{3},1)/2$, $k_\theta=2k_{\ty D}\sin(\theta/2) \approx k_{\ty D} \theta$ with $k_{\ty D}$ being the Dirac momentum and $w$ the hopping energy.  Estimates based on tight-binding models for AB bilayer graphene suggest that $w \approx 110$meV, 
however recent measurements suggest
that $w$ might be considerably smaller for some epitaxially grown
layers\cite{small_w}.

In the presence of a magnetic field it is convenient to work in the Landau gauge $\bm{A}=B(-y,0)$ and express the Hamiltonian in the
representation of the basis states $| L n \alpha y\ra$
where $L=1,2$ labels the layer, $n$ is the LL index, $\alpha=A,B$ stands for the sub-lattice, and $y$ is the guiding center coordinate.
The intra-layer part of the Hamiltonian is diagonal in $y$, however the $T_2$ and $T_3$ inter-layer hopping terms
change $y$ by $\pm \Delta$ where $\Delta=\sqrt{3}k_\theta \ell^2/2$.
In the presence of a finite $B$ the Hamiltonian therefore describes particles hopping on a
set of one-dimensional chains.
The Hamiltonian can be block-diagonalized in $y$ by grouping guiding centers separated by
integer multiples of $\Delta$ (see Supplementary Information).

The guiding center chains become periodic when $\Phi_0/\Phi$ is rational,
allowing a second wave vector to be introduced.
The corresponding basis functions are constructed by writing the $y$-guiding coordinate as $y=y_0+(m q + j)\Delta$
and Fourier transforming with respect to $m$.
The resulting magnetic Brillouin zone is:
\be
\{ (k_1,k_2)| 0<k_1=y_0/\ell^2<\Delta/\ell^2 , 0<k_2<2\pi/q\Delta \}.
\ee
The Hamiltonian matrix in this magnetic Bloch representation has dimension $4q$ times the number of Landau levels retained.

We numerically diagonalized the Hamiltonian for various twist angles accounting for inter-Landau level transitions (see Supplementary Information). Because $\alpha \propto \theta^{2}/B$ the sub-band structure
becomes more conspicuous as $\theta$ is reduced. On the other hand, the band structure for very small twist angles does not have simple Dirac character even at $B=0$\cite{moireBands}.   In Fig.\ref{fig:Esupport} we show the support of the spectrum for  the
intermediate case $\theta=2^\circ$.
As the magnetic field is increased (i.e. as $\alpha$ is decreased) all the gaps widen. The terminology of Landau level splitting is useful as
long as the single layer Landau levels do not overlap.
Mini-gaps as large as $10$meV open up within the $n=0$ Landau level for $B \approx 40$T.  When any one of the three
tunneling processes $T_j$ is present alone, the $n=0$ Landau level splits into two precisely degenerate components.  The relatively large
gap at the $\nu=0$ neutrality point which is present over a wide range of $\alpha$ in Fig.\ref{fig:Esupport} is a remnant of this behavior which
often remains when all three hopping processes are restored.

\begin{figure*}[h]
\includegraphics[width=0.8\linewidth]{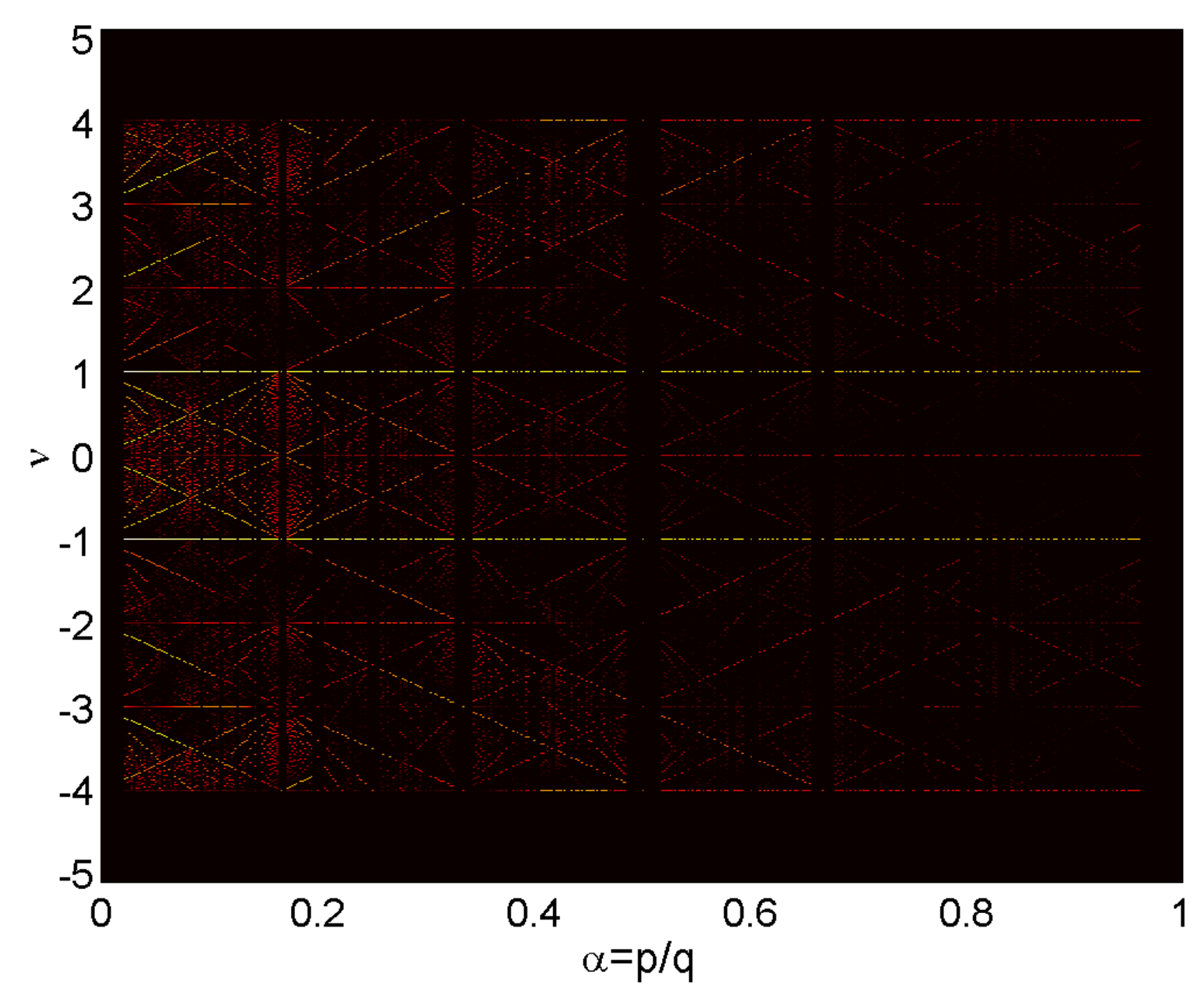}
\caption{{\bf Hall conductivity.}
A color plot for the energy gaps as a function of filling factor $\nu$ and $p/q$ for $\theta=2^\circ$ ($w=110$meV).
The color scale corresponds to $\log(1+\textrm{gap})$ in order
to magnify the smaller gaps. The conspicuous straight lines satisfy the Diophantine gap equation (\ref{diophantine}) and determine both $\sigma_{\ty H}$ and $s$.
The Hall conductivity corresponds to the intercept of the line with the y-axis whereas $s$ is given by its slope.}
\label{fig:gaps}
\end{figure*}

Since the pioneering work of Thouless {\em et al.}\cite{TKNN} it has
been understood that the Hall conductivity $\sigma_{\ty H}$ (in units of $e^2/h$) is a topological number that must be quantized
when the chemical potential lies in an energy gap.
Although the support of the spectrum as a function of field has a fractal structure, gaps in the spectra
can exist continuously over finite ranges of field.  The Landau level filling factors $\nu$ at which
gaps appear are characterized by two topological integers\cite{TKNN} which satisfy
\be
\nu = \sigma_{\ty H} + s \alpha     \label{diophantine}.
\ee
Here
\be
s = -\frac{\Omega}{A} \lp \frac{\partial N}{\partial \Omega} \rp_{B} = \frac{\Omega}{A} \frac{\partial^2 \cF}{\partial \mu \; \partial \Omega},  \label{s}
\ee
$A$ is the sample's area, $\Omega$ is the area of the unit cell, $N$ is the number of electrons in states below that gap, and $\cF$ is the grand canonical potential.
As a function of $\nu$ and $\alpha$, the Diophantine equation (\ref{diophantine}) has an infinite number of solutions:
$(s,\sigma_{\ty H}) = (s_0-mq , \sigma_0+mp)$ where $(s_0,\sigma_0)$ is some particular solution and $m$ is any integer.    
While there is a simple rule to determine $s$ is the classic Hofstadter problem\cite{TKNN} 
for the moir\'{e} butterfly $s$ and $\sigma_{\ty H}$ must be determined numerically; for eaxample by plotting the energy gaps
as a function of $\nu$ and $\alpha$.
The linear dependence assured by equation (\ref{diophantine}) allows a straightforward identification of
$\sigma_{\ty H}$ as the intercept of gap lines with the $\alpha=0$ axis
and of $s$ as the gap-line slope.
In Fig.\ref{fig:gaps} the energy gaps are plotted for $\theta=2^\circ$.

Using Fig.\ref{fig:gaps} we can identify the topological quantum numbers for
every gap in Fig.\ref{fig:Esupport}, as illustrated for some of the larger gaps that appear near
$\alpha \approx 0.3$. The integers depicted in Fig.\ref{fig:Esupport} specify the quantized Hall conductance in the
large gaps which appear between
$\nu=1$ and $\nu=-1$.
As the electronic density is varied the quantized Hall conductivity
follows the non-monotonic variation $+1,-1,+2,-2,+1,-1$.

As evident from equation (\ref{s}), in the case of bilayer graphene, the quantum number $s$ can be associated with the
chemical potential dependence of a rotational torque.  Measurement of this
electro-mechanical quantum number presents an interesting challenge to experiment.

Our theory has intriguing consequences for magneto-transport in double-layers grown using chemical vapor deposition (CVD).
This type of sample is polycrystalline in nature, characterized by graphene flakes of various sizes that are misoriented relative to one another.
A double-layer CVD grown structure will therefore be characterized not by a single twist angle but by a set of $\theta$'s. In the presence of a magnetic field the Hall conductivity of each
domain will depend on $B$ {\em and} on the particular twist angle of the domain.
Because different grains will in general have different Hall conductivities, chiral currents will flow
along most grain boundaries.

We note that the considerations presented here do not account for electron-electron interactions\cite{eeInteractions}.
As in the Hofstadter butterfly, fractional quantum Hall states with fractional charge and statistics are possible\cite{FQHE_read,FQHE_Allan}.
The large mini-gaps depicted in Fig.\ref{fig:Esupport} and the typical high mobilities of graphene multi-layers are favorable
for the experimental observation of these fractional states in exfoliated double-layer graphene samples.

We acknowledge a helpful conversation with Jo$\tilde{a}$o Lopes dos Santos, and thank Gene Mele for pointing out the difference between the moir\'e pattern and the $T(\bm{r})$ periodicity.
This work was supported by Welch Foundation grant F1473 and by the NSF-NRI SWAN program.

\newpage

\newpage

\newpage

\centerline{\Large \bf Supplementary Information}

\section{Hamiltonian}

To find the spectrum of the Hamiltonian given by equation (1) we represent it in terms of the basis states $| Ln\alpha y \ra$ where $L=1,2$ labels the layer, $n$ is the Landau level index,
$\alpha=A,B$ stands for the sub-lattice and $y$ is the guiding center coordinate.
The intra-layer part of the Hamiltonian is then
\be
h(\theta) = -\omega_c \sum_{Lny} \lp e^{-i\theta} \sqrt{n+1}|L n+1 A y\rangle\langle L n B y| + h.c.  \rp,
\ee
and the inter-layer part is
\bea
T = \sum_{nmy\alpha\beta} \lp T^{(0)} |2n\alpha y\rangle\langle 1m\beta y| + T^{(R)} |2n\alpha y+\Delta\rangle\langle 1m\beta y|   +
 T^{(L)} |2n\alpha y-\Delta\rangle\langle 1m\beta y|  \rp.      \label{T_LL}
\eea
Here $\Delta=\frac{\sqrt{3}}{2}k_\theta \ell^2$,
\bea
T^{(0)} &=& T_1 F_{nm}\lp \frac{\bm{q_1} \ell}{\sqrt{2}} \rp e^{-ik_\theta y}   \nonumber \\
T^{(R)} &=& T_2 F_{nm}\lp \frac{\bm{q_2} \ell}{\sqrt{2}} \rp e^{\frac{i}{4}k_\theta(2y-\Delta)}   \nonumber \\
T^{(L)} &=& T_3 F_{nm}\lp \frac{\bm{q_3} \ell}{\sqrt{2}} \rp e^{\frac{i}{4}k_\theta(2y+\Delta)},     \label{T_0RL}
\eea
(the $T_j$'s are defined by equation (3) in the main text) and
\be
F_{nm}(\bm{z}) = \sqrt{\frac{m!}{n!}} \lp -z_x+iz_y \rp^{n-m} e^{-\frac{z^2}{2}} \cL_m^{n-m}\lp z^2 \rp     \label{F}
\ee
for $n \geq m$ with $\cL$ being the associated Laguarre polynomial. For $n<m$ the function $F$ can be found using $F_{nm}(\bm{z}) = F_{mn}^\star(-\bm{z})$.

The commensurability condition $p/q=\Phi_0/\Omega_{\ty M} B$
is equivalent to the condition $k_\theta\Delta/2=2\pi p/q$ for which the hopping amplitudes (\ref{T_0RL}) repeat
their values when $y$ is shifted by $q\Delta$. As explained in the main text, this last periodicity together with the $y_0$ guiding center
coordinate define a magnetic Brillouin zone. For each momentum $\bm{k}=(k_1,k_2=y_0/\ell^2)$ in that zone
\bea
&& T(\bm{k}) = \sum_{nm\alpha\beta j}  \bigg[ T_j^{(0)} |2n\alpha j\rangle\langle 1m\beta j|
+ T_j^{(R)} |2n\alpha ,j+1\rangle\langle 1m\beta j|
+ T_j^{(L)} |2n\alpha j-1\rangle\langle 1m\beta j|  \bigg],      \label{T}
\eea
where $j=0,1, \ldots q-1$ ($j$ is defined modulo $q$ so that $|j=q\ra = |j=0\ra$), and
\bea
T_j^{(0)} &=& T_1 F_{nm}\lp \frac{\bm{q_1} \ell}{\sqrt{2}} \rp e^{-ik_\theta y_0} e^{-4\pi i \frac{p}{q}j}  \nonumber \\
T_j^{(R)} &=& T_2 F_{nm}\lp \frac{\bm{q_2} \ell}{\sqrt{2}} \rp e^{ik_2\Delta} e^{\frac{i}{2}k_\theta y_0} e^{i\pi\frac{p}{q}(2j-1)}   \nonumber \\
T_j^{(L)} &=& T_3 F_{nm}\lp \frac{\bm{q_3} \ell}{\sqrt{2}} \rp e^{-ik_2\Delta} e^{\frac{i}{2}k_\theta y_0} e^{i\pi\frac{p}{q}(2j+1)}.       \label{Ts}
\eea
Because $\lp k_\theta \ell  \rp^2 = 8\pi p/\sqrt{3}q $, the inter-layer Hamiltonian depends only on $p/q$.
In the absence of Landau level mixing the splitting of each Landau level into sub-bands is therefore determined only by $\alpha$.

Inter-Landau level transitions can significantly alter the electronic spectrum\cite{AllanLLmixing,klitzing2}.
In the absence of a magnetic field two momentum states are effectively coupled if their energy difference is of order of $E_{\Lambda}=max(v k_\theta,w)$ or less.
The same criteria holds also in the presence of a magnetic field. The $n=0$ Landau level therefore couples most strongly to
the $n_0 \approx (E_{\Lambda}/\omega_c)^2$ Landau level. In our calculations we retain $2n_0$ Landau levels in order to obtain
spectra that are accurate near the Dirac point. Comparing Fig.1 with results obtained
neglecting Landau level mixing (not shown) we find that, as expected, inter-Landau level hopping is increasingly
important as the magnetic field is reduced and as the energy is increased.

A naive approach to implement an energy cutoff would be to keep all states whose Landau level index is less
than $2n_0$. The problem with this approach is that it introduces a fake zero energy state irrespective of the size of $n_0$. We therefore make the energy cutoff
such that only one sub-lattice of the $n_0$ Landau level is retained. This approach shifts the numerical error to high energies, of the order of $\sqrt{n_0} \omega_c$.

\section{Moir\'e unit cell}

\begin{figure}[h]
\includegraphics[width=0.5\linewidth]{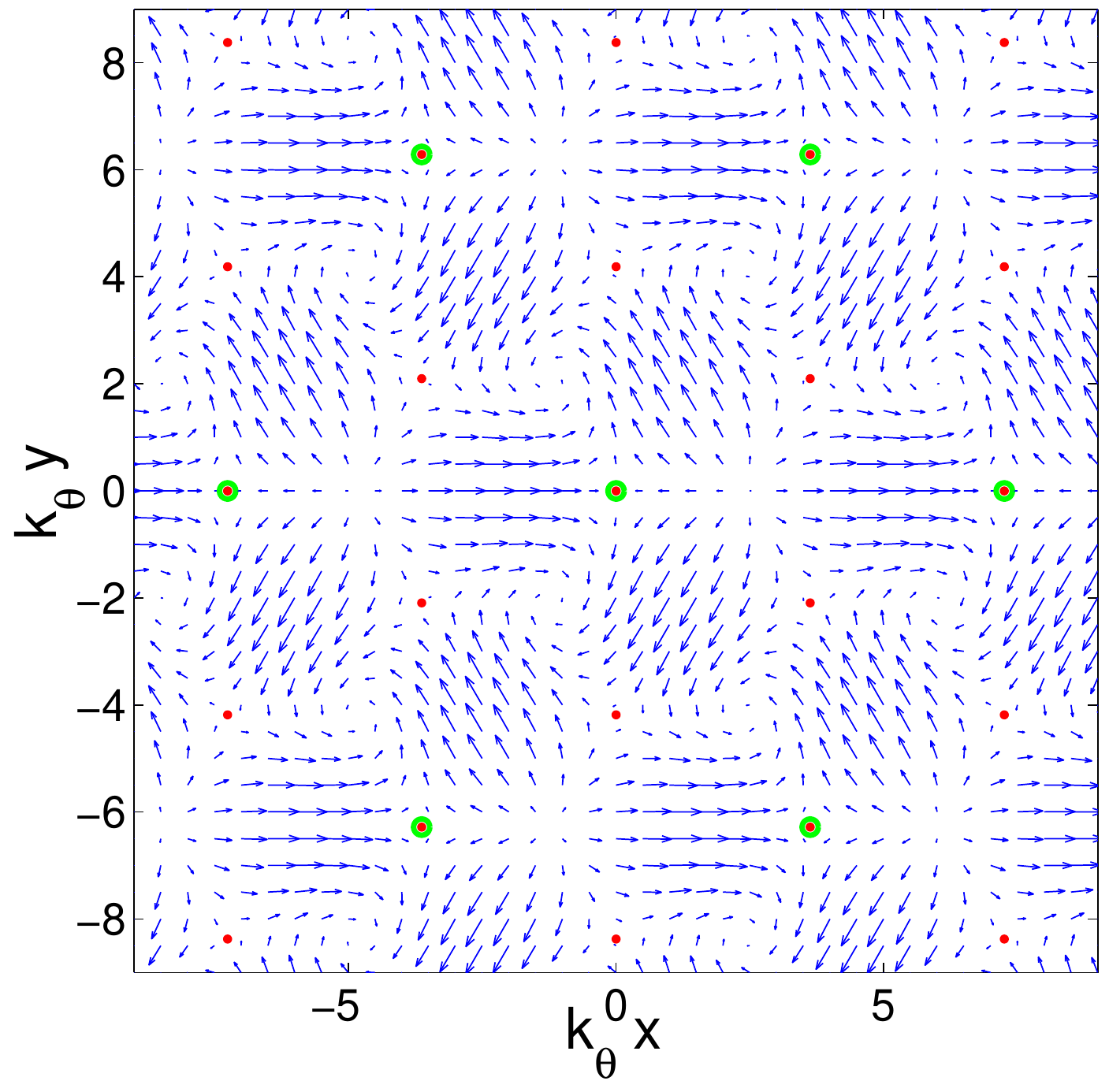}
\caption{
{\bf Moir\'e unit cell.} The AA entry of the hopping amplitude is plotted as a function of position (in units of $k_\theta^{-1}$). The space dependent eigenvalues of $H_{\ty T}$ lead
to the moir\'e pattern period marked by the red dots, however the spatial variation of the phase of $T$ results in a larger period denoted by the green circles. It is this latter periodicity that determines the moir\'e unit cell area $\Omega_{\ty M}$.}
\label{fig:moireUnitCell}
\end{figure}

The inter-layer hopping in twisted double layer graphene is akin in some respects to a periodic potential. In a real space representation it is described by the
$4 \times 4$ matrix (layer $\times$ sub-lattice)
\be
H_{\ty T} = \left(
  \begin{array}{cc}
    0 & T(\bm{r}) \\
    T^\dagger(\bm{r}) & 0 \\
  \end{array}
\right)        \label{TT}
\ee
where $T(\bm{r})$ is given by equation (2) in the main text.
The $AA$ entry of the $T$-matrix is depicted in Fig.\ref{fig:moireUnitCell} (similar figures are obtained for the other entries of $T$). The direction and size of an arrow at point
$\bm{r}$ correspond, respectively to the phase of $T_{\ty{AA}}(\bm{r})$ and to its magnitude.
The red dots mark the lattice associated with the moir\'e pattern whereas the green circles mark the lattice induced by $T(\bm{r})$.
Interestingly, the moir\'e unit cell area
\be
\Omega_{\ty M} = 16\pi^2/\sqrt{3}k_\theta^2
\ee
is six times larger than the area of the moir\'e pattern unit cell\cite{moireBands}.

\end{document}